\documentclass[epj]{svjour}

\usepackage{epsf}
\usepackage[T1]{fontenc}
\usepackage{amsmath}
\usepackage{amsfonts}
\usepackage{amssymb}
\usepackage{graphicx}
\usepackage{psfrag}

\newcommand{\beq}{\begin{equation}}
\newcommand{\eeq}{\end{equation}}
\newcommand{\beqn}{\begin{eqnarray}}
\newcommand{\eeqn}{\end{eqnarray}}

\begin{document}

\title{The effects of disorder and interactions on the Anderson transition
in doped Graphene}
\titlerunning{Disorder Effect in Graphene}
\author{Yun Song\thanks{\email{yunsong@bnu.edu.cn}}, Hongkang Song, and Shiping Feng }
\institute{Department of Physics, Beijing Normal University, Beijing
100875, China}
\date{\today}

\abstract{We undertake an exact numerical study of the effects of
disorder on the Anderson localization of electronic states in
graphene. Analyzing the scaling behaviors of inverse participation
ratio and geometrically averaged density of states, we find that
Anderson metal-insulator transition can be introduced by the
presence of quenched random disorder. In contrast with the
conventional picture of localization, four mobility edges can be
observed for the honeycomb lattice with specific disorder strength
and impurity concentration. Considering the screening effects of
interactions on disorder potentials, the experimental findings of
the scale enlarges of puddles can be explained by reviewing the
effects of both interactions and disorder.}

\maketitle

\section{Introduction}

As a two-dimensional (2D) allotrope of carbon
\cite{Novoselov,Zhang}, graphene holds great promise for replacing
conventional semiconductors on account of its unique electronic
properties. In graphene, the $\sigma-$band formed by the $sp^2$
hybridized orbitals determines the robustness of the honeycomb
structure, and the half-filled $\pi$-band provided with the
Dirac-like electronic excitations is responsible for the unusual
transport properties \cite{Neto}. It has been observed in some
experiments \cite{Bostwick,Adam,Tikhonenko,Martin,Chen,Tan} that
disorder has remarkable effects on the unusual electronic properties
of graphene, and thus there has been a great deal of interest in
recent years concerning the role of disorder in graphene.

Some experiments have confirmed the existence of metal-insulator
transition (MIT) in disordered graphene
\cite{Bostwick,Adam,Tikhonenko}, which suggests that the
one-parameter scaling theory \cite{Lee} may break down
\cite{Krapivsky}. In accordance with the prediction of Klein paradox
\cite{Katsnelson},
the Dirac fermions are found to be robust against disorder in the 2D
single valley Dirac model. In addition, the one-parameter scaling in
single-valley Dirac Hamiltonian shows that
\cite{Bardarson,NomuraRyu}, the logarithmic derivative
$d\mathrm{ln}\sigma/d\mathrm{ln}L=\beta(\sigma)$ is a positive
function and the scaling flow has no fixed point, indicating that
the system always scales towards a metal. This scenario is different
from what would be expected for the conventional 2D electron
systems, where all states are localized for arbitrary weak disorder
\cite{Lee}.

Recently,
some numerical methods have been adopted in the study of the
massless Dirac model with random potential
\cite{NomuraRyu,NomuraMacdonald,Ziegler,Guinea,Foster} or the
Anderson tight-binding model
\cite{Pereira08,Wu,ZhangLiu,WuLiu,Amini,Naumis,Xiong,Pershoguba,Schubert}.
It is already clear that the single valley Dirac approximation is
valid for graphene with weak long-range impurities, but the
intervalley scattering and chirality breaking scattering should be
considered in the presence of strong short-range disorder.
Therefore, the full tight-binding description of graphene with
Anderson disorder becomes a popular alternative model. Some very
recent simulations \cite{WuLiu,Amini,Naumis} have provided new
evidence for the existence of MIT in graphene with short-range
disorder by obtaining the mobility edges near the Fermi energy.
However, a contrary view has been supported by the transfer-matrix
approach \cite{Xiong} and kernel polynomial method \cite{Schubert}
that all states in the Anderson tight-binding model are localized
for arbitrary weak disorder. The possible reason for the
disagreement to arise is that different measures and scaling rules
are adopted in the above studies in an effort to distinguish
localized states from delocalized ones.

In this paper, we introduce a new approach to scale the inverse
participation ratio (IPR), which overcomes the unstable and
unreliable problems caused by the primitive scaling of IPR with the
negative first or second power of lattice size. We study the
tight-binding Hamiltonian of the honeycomb lattice with randomly
fluctuated disorder potentials, and an unique picture of Anderson
MIT has been obtained. As expected, we find that all electronic
states in graphene are localized by strong disorder. However, four
mobility edges can be observed when we decrease the disorder
strength until under a critical value $W_c$, and the electronic
states within two nearest neighbor mobility edges are entirely
extended or localized alternatively. Therefore, MIT can be
introduced by changing the carrier density to move the Fermi surface
across a certain mobility edge. Our results are in agreement with
the very recent experimental suggestions, where a transition from
localization to antilocalization has been achieved by decreasing the
carrier density in graphene samples \cite{Tikhonenko}.

On the other hand, the sources of the coexistence of electron and
hole puddles observed by the scannable single-electron transistor
(SET) have been proved to be both the substrate-induced structural
distortion as well as chemical doping. While the length scale of
density fluctuations, which is more than 150nm \cite{Martin}, is
extremely larger than the disorder length scale introduced by the
short-range scatters. In this paper, we show that the unexpected
large disorder length scale observed by SET can be explained by the
screening effect of interactions on disorder, suggesting that the
electron-electron interactions should be taken into account.
Besides, the delocalization effect of interactions through screening
disorder potentials can also introduce the zero-bias anomaly at
Fermi surface \cite{Efros-a,Efros-b,Altshuler}.

In this paper we address the following issues related to the effect
of disorder in graphene: (1) Whether there exists the mobility edge
to separate the localized states from the delocalized states in
disordered graphene? A distinctive scenario of Anderson MIT with
four mobility edges have been observed in graphene with quenched
random disorder; (2) How to measure the localization length of a
localized electronic state by IPR? We present a new method to set
the lattice size scaling of IPR, which can give the localization
length of electronic state precisely; (3) What is the role of
interactions in disordered graphene? The screening effect of the
short-rang interactions on disorder potentials is found to have
strong affection for the scale enlarges of puddles.

The paper is organized as follows: In section.~\ref{Model} we
present the fully tight-binding description of the disordered
graphene. In section.~\ref{SCIPR} and \ref{SCGADOS}, we discuss the
lattice size scaling of the IPR and the geometrically averaged
density of states (GADOS) respectively, and show the scaling method
relevant for the subsequent discussions. Next we show our findings
regarding the static disorder in graphene in
secrion.~\ref{Result_dis}. In addition, the screening effects of the
electron-electron interactions on the disorder potentials are
presented in section.~\ref{Result_int}. The principal findings of
this paper are summarized in section.~\ref{conclusion}.

\section{Theory}

\subsection{The Anderson Tight-banding Model}
\label{Model}

The Anderson tight-banding Hamiltonian \cite{Anderson,Evers} for the
disordered graphene can be expressed as
\begin{equation}
H_{AM}=-t\sum_{\langle ij\rangle}c_{i}^{\dagger}c_{j} +\sum_{i}
\epsilon_i c_{i}^{\dagger}c_{i}, \label{Eq.Ham}
\end{equation}
where $c_{i}$ ($c_{i}^{\dagger}$) are annihilation (creation)
operators of a fermion at site $i$, $\epsilon_i$ represent the
on-site disorder energies, and $t$ denote the hoping integrals
between nearest neighbor (NN) sites, which are set to be unit in our
calculations. We take the exact numerical approach to find all
eigenenergies and eigenfunctions of the Hamiltonian given by
Eq.~(\ref{Eq.Ham}) for a finite hexagonal lattice with the periodic
boundary conditions. Besides, the numerical calculations are
performed with regard to the following three different disordered
cases.

Firstly, in the presence of quenched random disorder, we introduce
the box distributed Anderson disorder, and the on-site disorder
energies $\epsilon_i$ in Eq.~(\ref{Eq.Ham}) are assumed to be
independent random variables distributed between $-W/2$ and $W/2$.
Therefore, the probability distribution of $\epsilon_i$ is given by
\begin{equation}
P(\epsilon)=W^{-1}\Theta(W/2-|\epsilon|),
\end{equation}
and the disorder strength is parameterized by the width $W$.

Secondly, to take into account the influences of the impurity
potentials to the adjacent sites \cite{ZhangLiu}, the correlation
length of disorder $\zeta$ is introduced and the on-site disorder
energies $ \epsilon_i$ in Eq.~(\ref{Eq.Ham}) can be replaced by the
correlated disorder energy $\widetilde{\epsilon}_i$ as
\begin{equation}
 \widetilde{\epsilon}_i=\sum_{l=1}^{N}\epsilon_l
exp(-|\textbf{r}_i-\textbf{r}_l|^2/\zeta^2). \label{Eq.Cor}
\end{equation}
Where $\epsilon_l$ represent the "uncorrelated" disorder energies of
the impurities at sites $l$, which have been defined in the first
case.

Finally, in order to generalize this Hamiltonian by including
electron-electron correlations, one must add an interaction term
$H_{int}$ onto the Anderson tight-binding model:
\begin{equation}
H=H_{AM}+H_{int},
\end{equation}
with
\begin{equation}
H_{int}=U\sum_in_{i\uparrow}n_{i\downarrow}+\sum_{ij}V_{ij}n_in_j.
\end{equation}
Here $n_{i\sigma}=c^{\dagger}_{i\sigma}c_{i\sigma}$ is the local
charge-density operator with $\sigma=\uparrow,\downarrow$, $U$
represent the on-site interactions, and $V_{ij}$ denote the
non-local interactions between electrons at sites $i$ and $j$. To
compare with the experimental finding about the enlarges of puddles
\cite{Martin}, here we consider a case of the binary alloy disorder,
in which the probability distribution function of the on-site
disorder energies is given by
\begin{equation}
p(\epsilon_i)=x\delta(\epsilon_i-W)+(1-x)\delta(\epsilon_i),
\end{equation}
where $x$ is the fraction of the lattice sites with energies
$\epsilon_i=W$, and $W$ represents the on-site energy splitting.

\subsection{The Lattice Size Scaling of IPR}
\label{SCIPR}

One of the widely used quantities to measure the Anderson
localization of electronic states in disordered systems is called
IPR, which is defined as \cite{Thouless,Wegner}
\begin{equation}
I_2(E_n)=\sum_{i=1}^N |\Psi_n(r_i)|^4,\label{IPR}
\end{equation}
where $E_n$ ($n$=1,...,N) and $\Psi_n(r_i)$ are the eigenenergies
and eigenfunctions of a disordered finite lattice with $N$ sites.
Since the eigenenergies of a finite lattice are always discrete, it
is more convenience to introduce a continuous energy-dependent IPR
as
\begin{equation}
I_2(\omega, N)=\frac{\sum_{i=1}^N
\Theta(\frac{\gamma}{2}-|\omega-E_n|)I_2(E_n)} {\sum_{i=1}^N
\Theta(\frac{\gamma}{2}-|\omega-E_n|)},
\end{equation}
where $\omega$ represents energy, and  $\gamma$ is a very small
energy bin for the average. Since IPR of extended states are
proportion to $1/L^d$ and go to zero in the infinite lattice limit
$L\rightarrow\infty$, it is not difficult to identify extended
states by doing lattice size scaling of IPR. Here $d$ denotes the
dimension of the system, and $L$ represents the lattice size with
$N=L^d$. However, the lattice size scaling of IPR for a localized
state is found to be more complicated than expected
\cite{Brndiar,Cuevas,Zekri,Song}.

The wave function of Anderson localized states decays exponentially
as $|\psi(r)|\sim exp(-|r-r_0|/\xi_{loc})$, where $\xi_{loc}$ is the
localization length. When the lattice size $L$ is much larger than
the localization length, IPR is a size-independent constant $I^0_2$,
and the localization length can be obtained easily by
$\xi_{loc}=(I^0_2)^{-1/d}$ \cite{Song}. However, we can not always
perform calculations satisfying the condition $L\gg\xi_{loc}$ on
account of the limited capacity of the numerical calculations. When
the lattice sizes are comparable to or even smaller than the
localization length $\xi_{loc}$, $L$ becomes a variable to the
function of IPR. Later a simple lattice size scaling formula
$1/L^{\alpha}$ has been suggested. Unfortunately, this IPR scaling
can mistakenly regard a localized state as an extended state in some
cases. For example, the exact scaling formula of IPR in
one-dimensional (1D) systems with box distributed disorder has been
obtained as \cite{Song}
\begin{equation}
I_2(\omega,L)=I_2(\omega,\infty)\coth(L/2\xi_{loc}(\omega)),
\label{1DIPR}
\end{equation}
with $\xi_{loc}(\omega)=I_2^{-1}(\omega,\infty)$. It is obvious that
the simple formula $1/L^{\alpha}$ is not a proper approximation in
considerable large lattice size region to scale IPR. Certainly,
there are also exceptions for the conditions with $L\gg\xi_{loc}$
and $L\ll\xi_{loc}$, where we can choice $\alpha=0$ or $\alpha=2$
respectively.

\begin{figure}[tb]
\centerline{\epsfxsize=2.75in\epsfbox{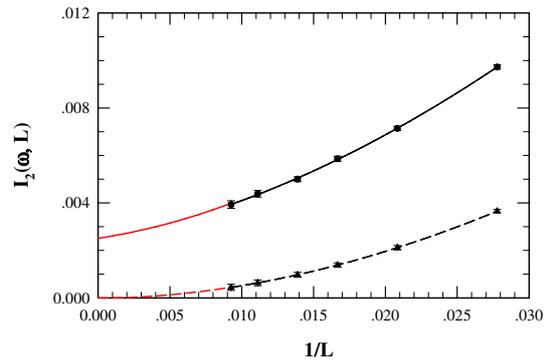}}
\caption{(color online) The lattice size scaling of IPR $I_2(\omega,
L)$ for a localized (circles) and an extended (triangles) state. The
black lines are the fitting curves obtained by the polynomial
formula shown in Eq.~(\ref{Taylor}), and the red lines show the
extensions of the fitting curves to the infinite lattice limit
$L\rightarrow\infty$.} \label{fig:IPR}
\end{figure}

Unlike the 1D disordered system, the exact expression for IPR has
not been obtained for the 2D finite systems. Therefore, we introduce
a Taylor series to scale IPR by
\begin{equation}
I_2(\omega, L)=I_2(\omega,\infty) +\frac{a_1(\omega)}{L}
+\frac{a_2(\omega)}{L^2} +\frac{a_3(\omega)}{L^3} +...,
\label{Taylor}
\end{equation}
where $a_n(\omega)$ is the $n$-th Taylor parameter. The minimum
radius of convergence  is found to be $R_{min}=0.1$ in our studies,
and to make all Taylor series adopted convergent, we have to do
calculations when lattices meet the condition of $L>1/R_{min}=10$.
For convenience, we employ a polynomial formula consist of the front
five terms in Eq.~(\ref{Taylor}) since the contributions of
high-order terms are negligible. As shown in Fig.~\ref{fig:IPR}, the
new fitting is reasonable and pratical since the intercept of an
extended state is found to be zero or even a very small negative
number, whereas the fitting curve of a localized state has a
positive intercept in the infinite-lattice limit. We have checked
the above fitting method by the disordered cubic lattice
\cite{Yang}, and good agreements have been achieved with the
accepted scaling theories \cite{Lee}. Especially, we find that the
fitting curves of extended states can be approximated by
$I_2(\omega, L)\propto 1/L^2$, in according with the prediction of
some other theories. On the other hand, when the electronic state is
localized, the five parameters are all found to have non-zero
values, but the low-order terms play the main roles. As show by the
solid line in Fig.~\ref{fig:IPR}, there exists a finite intercept
$I_2(\omega, \infty)$ for a localized state, and its localization
length can be obtained by
\begin{equation}
\xi_{loc}(\omega)=\frac{1}{\sqrt{I_2(\omega, \infty)}}.
\end{equation}

To sum up, the new scaling method of IPR has the advantage in
distinguishing explicitly the localized states with extended states,
and in addition we can acquire the localization length precisely by
the intercept obtained in the infinite-size limit.

\subsection{Measure Anderson Localization by GADOS}
\label{SCGADOS}

It has been proposed that Anderson localization can be measured by
an order parameter $\rho_g(\omega)$, called GADOS, which is obtained
by geometrically averaging the local DOS (LDOS) of each site as
\begin{equation}
\rho_g(\omega)=\frac{1}{N_s}\sum^{N_s}_{m=1}\left[\prod^N_i
\rho(r_i,\omega)\right]^{1/N},
\end{equation}
with
\begin{equation}
\rho(r_i,\omega)=\frac{\sum_{i=1}^N
\Theta(\frac{\gamma}{2}-|\omega-E_n|)\rho(r_i,E_n)} {\sum_{i=1}^N
\Theta(\frac{\gamma}{2}-|\omega-E_n|)}.
\end{equation}
Where $\rho(r_i,E_n)=|\Psi_n(r_i)|^2$ is the LDOS at site $i$ for
the $n$-th eigenstate $\Psi_n$, $N_s$ represents the number of
disorder configurations to be averaged, and the energy bin $\gamma$
plays the same role as in Sec.~\ref{SCIPR}.

For the infinite-dimensional disordered system, it has been defined
that Anderson transition happens when $\rho_g(\epsilon_F)$ vanishes
completely at Fermi surface $\epsilon_F$. This criterion has been
introduced to investigate the competition between Anderson
transition and Mott MIT in the infinite-dimensional systems within
the dynamical mean-field theory \cite{Dobrosavljevic,Byczuk}. While,
in the numerical calculation for a finite lattice of low-dimensional
disordered system, a finite energy bin $\gamma$ has to be
introduced. As a result, GADOS is a function of lattice size, and
its lattice size scaling should be introduced to detect the Anderson
MIT.

In Ref.~\cite{Song}, we have examined the lattice size scale of
GADOS and found that, for any nonzero energy bin $\gamma$, $\rho_g$
decays exponentially with the increasing of lattice size for a
localized electronic state. On the contrary, there is no significant
variation of $\rho_g$ with the increasing of lattice size for a
delocalized state. Therefore, it is not difficult to distinguish
localized states from extended states by the lattice size scaling of
GADOS. In this paper, we use GADOS scaling to check our results
obtained by IPR scaling about whether the electronic states are
localized or not, and good agreements have been achieved.

\begin{figure}[tb]
\centerline{\epsfxsize=2.7in\epsfbox{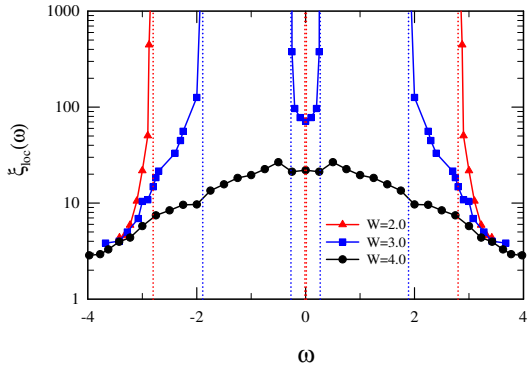}}
\caption{(color online) The energy dependencies of the localization
length $\xi_{loc}(\omega)$ of graphene with Anderson disorder. The
box distribution is adopted for the on-site energies of every sites
with disorder strength $W$=2.0 (red triangles), 3.0 (blue squares),
and 4.0 (black circles). The dotted lines denote the mobility
edges.} \label{fig:MIT}
\end{figure}

\section{Localization in disordered graphene}
\label{Result_dis}

The scaling theories of localization have proved that all electronic
states should be localized in the conventional 2D systems with
arbitrary weak disorder \cite{Lee}. Comparing with the parabolic DOS
of conventional 2D system, graphene has a special band structure
with zero weight at Dirac points.  Applying the Dirac model with
random potentials to the weakly disordered graphene, some studies
have confirmed the existence of the antilocalizatized electronic
states \cite{NomuraMacdonald,NomuraRyu} because of the insensitivity
of the Dirac fermions to the disordered external electrostatic
potentials. On the other hand, strong disorder can manifestly
enhance the DOS at Dirac points and strongly affect the electronic
properties. For this reason, our study will concentrate firstly on
the evolutions of electronic states in the whole energy band with
the increasing of disorder strength.

We study the Anderson tight-binding model on a hexagonal lattice by
exact numerical simulations, where the largest lattices could be
120$\times$120. Applying the IPR scaling method introduced in
Sec.~\ref{SCIPR}, the localization lengths of localized electronic
states are obtained for different disorder strength $W$. As shown in
Fig.~\ref{fig:MIT}, it is obvious that disordered graphene has a
quite different scenario of Anderson localization than the
conventional 2D systems with disorder. When disordered strength
$W=2$ or 3, we find energy regions consist of the localized
electronic states separated by the mobility edges from the regions
of extended ones, where the physical observable is the finite
localization lengths of localized states. Close to the top and
bottom of the whole energy band, we obtain two symmetrical mobility
edges at $\omega=\pm\omega_{c_1}$. In addition, two mobility edges
appear near the Fermi surface ($\epsilon_F=0$) with
$\omega=\pm\omega_{c_2}$($\omega_{c_2}<\omega_{c_1}$), suggesting
the electronic states around the Fermi surface are also easier to be
localized by the scattering with the disorder potentials.
Furthermore, all electronic states can be localized when the
disorder strength is increased larger than $W_{c}^{\prime}=3.10$.

The region close to the Dirac points is very important, and people
show great interests in the possible existence of the antilocalized
electronic states at there. However, the scaling results are
affected magnificently by the relative big errors for IPR since the
DOS is quite low in the vicinity of Dirac points. In addition, it is
well known that to distinguish an extended state from a weakly
localized state with large localization length is very difficult. To
make our finding more convincing, we also use the scaling of GADOS
to measure Anderson localization in disordered graphene. We find
very good agreements between the scalings of GADOS and IPR.

\begin{figure}[tb]
\centerline{\epsfxsize=2.75in\epsfbox{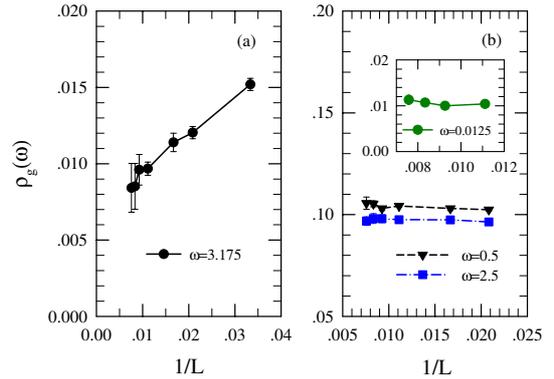}}
\caption{(color online) The lattice size scaling of GADOS of
electronic states with different energies: (a) $\omega=3.175$ (black
circles), and (b) $\omega=$0.5 (black triangles) and 2.5 (blue
squares). Insert: the scaling of GADOS close to the Dirac point with
$\omega=0.0125$ (green circles). The disorder strength of Anderson
disorder is $W$=2.0.} \label{fig:GADOS}
\end{figure}

In Fig.~\ref{fig:GADOS}, we show the lattice size dependencies of
GADOS for some representative energies when $W=2$. In
Fig.~\ref{fig:GADOS}(a), GADOS of $\omega=3.25$ increases manifestly
with the increasing of $1/L$, suggesting the state is localized. On
the contrary, GADOS of the other two extended states with energy
$\omega=0.5$ and $2.5$ are both lattice size independent as shown in
Fig.~\ref{fig:GADOS}(b).  When $\omega=0.0125$, we have performed
the scaling of GADOS very carefully and precisely by averaging more
than two hundreds of disorder configurations for lattice with
$L=120$. As shown by the insert in Fig.~\ref{fig:GADOS}(b), the
lattice size independent behavior of GADOS indicates clearly that
the electronic states quite close to the Dirac points are  still
extended, in accordance with the result obtained by the IPR scaling.
Our result suggests that the prediction of single valley model, i.e.
Dirac Fermions can not be localized by disorder, is reasonable in
weakly disordered graphene since no intervalley scattering can be
stimulated by very weak disorder. In addition, it is obvious that
the Anderson MIT can be introduced by only changing the carrier
density to move the Fermi surface from extended region to localized
region with fixed disorder strength. Recently, a transition between
localization and antilocalization has been observed experimentally
in graphene when the carrier density is decreased \cite{Tikhonenko}.
Therefore, our result about Anderson localization with four mobility
edges help clarify this experimental observation of MIT in graphene.
In addition, our findings suggest that the delocalization is closely
related to the linear dispersion near the Dirac points which may
suppress the inter-valley scattering. And further investigations are
to be done in the future study.

\begin{figure}[tb]
\centerline{\epsfxsize=2.9in\epsfbox{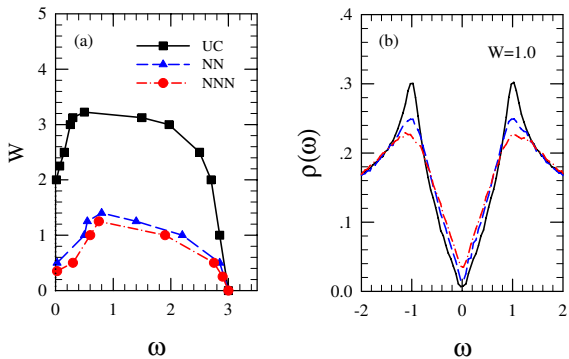}}
\caption{(color online) (a) The energy dependencies of the critical
disorder strength $W_c(\omega)$ of graphene with correlated
disorder; (b) The effects of the correlations of disorder on DOS
around Fermi surface $\omega=0$. The solid (black), dashed (blue),
and dashed-dotted (red) lines represent respectively the results of
point defects, NN scatters and both NN and NNN scatters. }
\label{fig:NNN}
\end{figure}

The electrostatic coulomb potentials of the surface absorptions or
the adatoms in the substrate for graphene should be described by
Eq.~(\ref{Eq.Cor}), in which the correlations between the disorder
potentials of individual sites are considered. Owing to the
screening effect of electrons, the long-range correlations of
disorder could be neglected. Here we name the uncorrelated and
short-range correlated disorder as point defects and short-range
scatters respectively. In Fig.~\ref{fig:NNN}, we show the effects of
the nearest-neighbor (NN) and next NN (NNN) scatters on the
localization of electronic states. The correlation length $\zeta$ is
set as $1.5a$, which is the same as the parameters chosen in
Ref.~\cite{Wakabayashi}, and the blue dashed and red dash-dotted
lines reveal the energy dependence of the critical disorder strength
$W_c(\omega)$ for graphene with the NN scatters only and both NN and
NNN scatters, respectively. For the convenience of comparison, we
also show the results of uncorrelated point defects (black solid
line). It is obvious that the picture of Anderson MIT for
short-range scatters is the same as that of the point defects.
While, the short-range correlations of disorder have strong effects
to localize the electronic states. As to why the critical strengthes
of disorder $W_c$ for localizing the states are greater for
uncorrelated disorder than the correlated ones, the effective
disorder strength is actually enhanced a lot as we add the
influences of the impurity potentials of the adjacent sites onto the
uncorrelated on-site energy of a certain site directly. As a result,
the localization of electronic states is enhanced accordingly when
the re-scaling is not introduced to reduce the original disorder
strength.

\section{The interactions and electron-hole puddles in disordered graphene }
\label{Result_int}

\begin{figure}[tb]
\centerline{\epsfxsize=2.75in\epsfbox{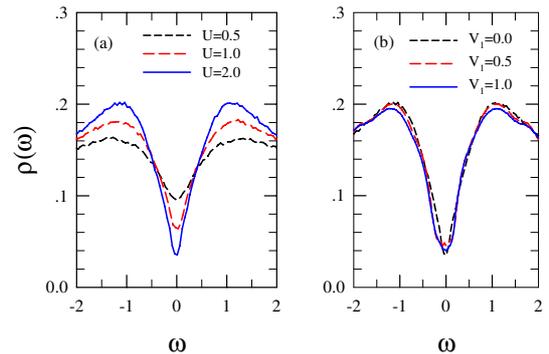}}
\caption{(color online) The evolutions of DOS of graphene with
disorder strength $W=4$ and short-range interactions: (a) there are
only on-site interactions $U=$0.5 (black dashed line), 1.0 (red
dashed line), and 2.0 (blue solid line), and (b) the on-site
interactions are fixed as $U=2.0$, and the NN interactions are given
as $V_1=0.0$ (black dashed line), 0.5 (red dashed line), and 1.0
(blue solid line).} \label{fig:DOS}
\end{figure}

In pure graphene, the influence of short-range interactions on
electronic properties has been investigated by the dynamical
mean-field theory (DMFT) \cite{Jafari}. The Dirac sea state is found
to remain stable against local many-body interactions since the
interactions in graphene are much smaller than the Mott critical
value $U_c=13.3t$ \cite{Jafari}. In the present of disorder, the
combine effects of interactions and disorder will have strong
influences on the LDOS and also the localization of electronic
states. It has been proved recently by the statistic DMFT
calculations \cite{Song} that the Hartree-Fock approximation (HFA)
could give reasonable results for the conventional 2D
Anderson-Hubbard model when the interactions $U$ are smaller than
the energy bandwidth $D$. Based on the predictions for the many-body
effect in graphene given by Ref.~\cite{Neto}, we study the
Anderson-Hubbard mode with weak interactions in the region of
$U=0.5t$ to $4.0t$ within HFA.

In conventional 2D disordered systems, the zero-bias anomaly at
Fermi surface \cite{Efros-a,Efros-b,Altshuler} arises from the
interplay between disorder and interactions, which indicates the
delocalization effects of interactions on localized electronic
states \cite{Song2}. In the same manner, we observe the DOS of
disordered graphene at Fermi surface to show the delocalization
effects of the short-range interactions. Firstly, we only consider
the effects of the on-site interactions $U$, and as shown in
Fig.~\ref{fig:DOS}(a), significant decreasing of the DOS around the
Fermi surface appears when we increase $U$ from 0.5 to 4.0. On the
contrary, the long-rang interactions are found to have less effect
to the zero-bias anomaly at Fermi surface. In Fig.~\ref{fig:DOS}(b),
we plot the DOS for graphene with different NN interactions and
fixed $U$, and it is obvious that even the effects of NN
interactions are still weak and negligible. Therefore, we predict
that the on-site interaction is the key point in the study of the
screening effects of interactions in disordered graphene. The local
effect of interactions can be understood by calculating the screened
potential $\upsilon_i=\epsilon_i(1-U\chi_{ii})$, where the local
charge susceptibility is defined as $\chi_{ii}=-\partial
n_i/\partial \epsilon_i$. It is obvious that the screening effect in
localized phase is expected to be less than in metallic phase since
$\chi_{ii}$ is restrained for site with small LDOS at Fermi level.
We find in the paramagnetic phase that the on-site interactions have
strong effects to delocalize the electronic states by increasing the
localization lengths accordingly.

\begin{figure}[tb]
\centerline{\epsfxsize=3.5in\epsfbox{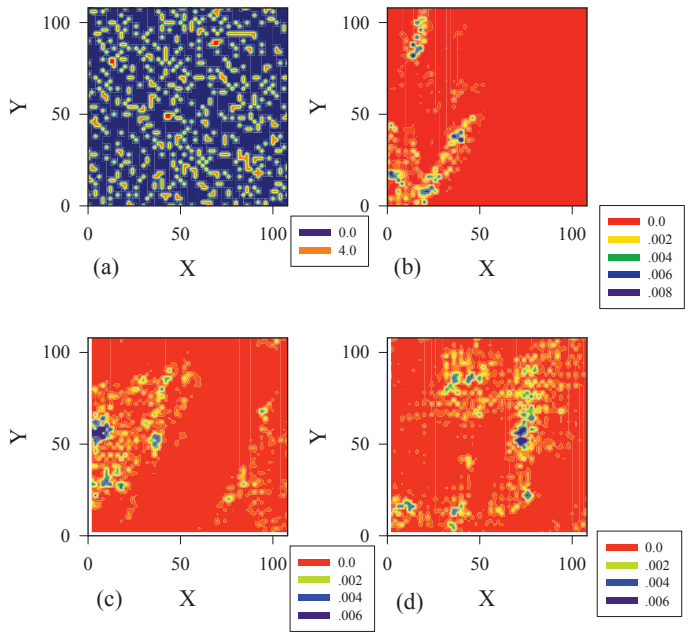}}
\caption{(color online) (a) The disorder configuration of an
108$\times 108$ lattice with binary disorder. The on-site energy
splitting is $W=4$; The corresponding LDOS of an eigenstate right at
the Fermi level for different on-site interactions: (b) $U=0.0$, (c)
2, and (d) 4.0. The systems are all half-filled with impurity
concentration $x=0.2$.} \label{fig:LDOS_HF}
\end{figure}

\begin{figure}[tb]
\centerline{\epsfxsize=3.5in\epsfbox{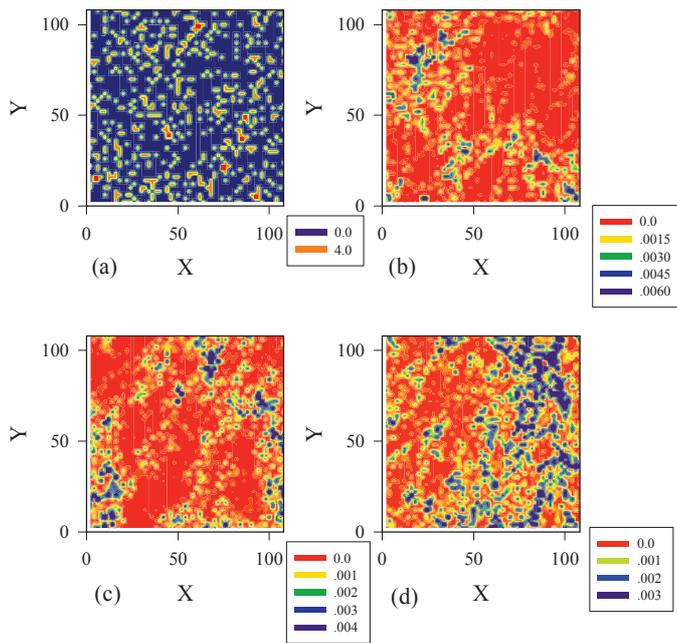}}
\caption{(color online) The disorder configuration and the
corresponding LDOS are shown for the same parameters as in
Fig.~\ref{fig:LDOS_HF}, except that the systems are all
quarter-filled.} \label{fig:LDOS}
\end{figure}

Since The LDOS at Fermi surface $\epsilon_F$ is detectable, to
compare with the experimental results, we calculate the evolution of
$\rho(r_i, \epsilon_F)$ with the on-site interactions $U$ for a
particular binary disordered configuration. As shown in
Fig.~\ref{fig:LDOS_HF} and Fig.~\ref{fig:LDOS}, the length scale of
density variations of LDOS for $U=0.0$ is much smaller than that for
$U=2.0$ and 4.0. Compared with the quarter-filling case, the
delocalization effects of interactions are not remarkable when the
system is half-filled. It is obvious that the states at Dirac points
are strongly localized at half-filling with $W=4$. This further
proves the results we obtained in the previous section from a
different perspective. The regions of electron-rich and hole-rich
puddles have been observed by the experiment of using scanning
single-electron transistor to map the LDOS of graphene sheet
\cite{Martin}. The smallest length scale on which density variations
is observed roughly 150nm, which is apparently significantly larger
than the intrinsic disorder length scale as approximately 30nm in
the graphene samples. There arises the question of why there exists
such a big difference between the length scales of the LDOS and
intrinsic impurities. Our results predict that the interactions play
an important role in the electron-hole puddles observed by
experiment, where the screening effects of the on-site interactions
on disorder potentials can enlarge the length scale of LDOS
significantly.

\section{Conclusions}
\label{conclusion}

We have studied numerically the Anderson tight-banding model of the
finite hexagonal lattices and found that, unlike the conventional 2D
disordered systems, there is a unique picture of the Anderson
localization with four mobility edges in disordered graphene. We
predict that Anderson MIT can be achieved in graphene by changing
the carrier density to make a move of the Fermi surface across the
mobility edges. In addition, we have found that the length scale of
LDOS is considerably enhanced by the on-site interactions when the
screening effects of interactions on the disorder potentials are
also taken into consideration.

Besides, the measure of Anderson localization by IPR has also been
discussed, and a polynomial formula has been introduced for the
lattice size scaling of IPR. As a result, precise localization
length of the localized electronic state can be obtained by the
intercept found in the infinite-size limit.

\section*{Acknowledgments}
We are grateful to W. A. Atkinson for reading the manuscript and
valuable comments. The work was supported by the project-sponsored
by SRF for ROCS, SEM, the NSFC of China, under Grant Nos. 10974018
and 10774015, and the National Basic Research Program of China
(Grant Nos. 2011CBA00108 and 2007CB925004).

\end{document}